# Cross-correlated quantum thermometry using diamond containing dual-defect centers

*Madhav Gupta[a], Tongtong Zhang[a], Lambert Yeung[b], Jiahua Zhang[a], Yayin Tan[a], Yau Chuen Yiu[a], Shuxiang Zhang[a,e], Qi Wang[c], Zhongqiang Wang[c], Zhiqin Chu\*[a,d]*

[a] Department of Electrical and Electronic Engineering, The University of Hong Kong, Pokfulam Road, Hong Kong SAR, China
[b] 6C Tech Ltd. Hong Kong SAR, China
[c] Dongguan Institute of Opto-Electronics, Peking University, Guangdong, China
[d] School of Biomedical Sciences, The University of Hong Kong, Pokfulam Road, Hong Kong SAR, China
[e] Sichuan University, China

E-mail: zqchu@eee.hku.hk

**Keywords**: Nitrogen-Vacancy Center, Diamond, Optical thermometry, Cross-validation

**ABSTRACT**: The contactless temperature measurement at micro/nanoscale is vital to a broad range of fields in modern science and technology. The nitrogen vacancy (NV) center, a kind of diamond defect with unique spin-dependent photoluminescence, has been recognized as one of the most promising nanothermometers. However, this quantum thermometry technique has been prone to a number of possible perturbations, which will unavoidably degrade its actual temperature sensitivity. Here, for the first time, we have developed a cross-validated optical thermometry method using a bulk diamond sample containing both NV centers and silicon vacancy (SiV) centers, achieving a sensitivity of 22 mK/√Hz and 155 mK/√Hz respectively. Particularly, the latter has been intrinsically immune to those influencing perturbations for the NV-based quantum thermometry, hence serving as a real-time cross validation system. As a proof-of-concept demonstration, we have shown a trustworthy temperature measurement under the influence of varying magnetic fields, which is a common artefact present in practical systems. This multi-modality approach allows a synchronized cross-validation of the measured temperature, which is required for micro/nanoscale quantum thermometry in complicated environments such as a living cell.

## 1. Introduction

Performing high-resolution thermometry with nanoscale spatial resolution is crucial in studying multiple processes in diverse fields such as electronics, material science, and cell biology[1-5]. For example, multiple physiological properties in thermal biology are directly revealed by measuring the temperature in biological systems[6]. There are growing interests in developing non-contact luminescent thermometry techniques to explore the measurements of temperature at single cell level[6,7]. Among the different approaches, the optically addressable nitrogen vacancy (NV) center in diamond nanoparticles remains one of the most-studied and well-adopted high-sensitive nanothermometers[8]. The NV center uses a magnetic spin transition whose resonance frequency is sensitive to thermally induced lattice expansion. This spin-based method has exceptionally high sensitivity, and has been shown to detect temperature variations down to 9 mK/√Hz in pure diamond and 200 mK/√Hz in nanodiamonds with a spatial resoultion of 200nm[7]. Several exciting applications of this method have been demonstrated, such as nanoscale thermometry in electronic devices[9], living cells[7], and C. elegans worms[10].

However, this spin-based method has been found susceptible to multiple artefacts such as fluctuating magnetic fields[11], microwave heating [12-13], and uncontrolled movements[8,14]. In addition, the required microwave irradiations lead to considerable residual heating effects and might not be suitable for biological samples such as neuron cells[15]. Due to these practical implementation issues, several microwave-free all-optical approaches have been demonstrated using a series of diamond defects such as NV[16], silicon vacancy (SiV) [17-18], germanium vacancy (GeV)[19], and tin vacancy (SnV)[20-21] centers. For instance, temperature measurement within the biological transparency window by the SiV defect in diamonds has been recently reported[18]. Here the temperature sensitivity reached √360 mk/Hz in bulk diamond and 521 mk/Hz in 200 nm nanodiamonds at room temperature. While the all-optical method has been shown to be relatively immune to those measurement artefacts, it suffers from relatively poor sensitivity as compared to the spin-based approach[8]. Notably, most of the existing studies have relied on one single type of defect centers, and only a few recent studies[22-23] have started to demonstrate the case of deploying dual-defects simultaneously in one go.

For reliable temperature measurements in practical applications, it is highly desirable to develop a multi-modality thermometer allowing cross-validation of each individual readout method[5]. Recently, multi-mode temperature measurements have been demonstrated, e.g., Alkahtani et al. reported the simultaneous use of fluorescent nanodiamonds (FNDs) and lanthanide ion-doped upconverting nanoparticles (UCNPs) in fertilized bovine embryos[22]. They injected both FNDs and UCNPs into the embryos, performing Optically Detected Magnetic Resonance (ODMR) measurements for the FNDs, and the fluorescence spectral measurements for the UCNPs aimed at increased measurement confidence. Despite its significance, this dual-mode thermometry has actually consisted of 2 physically separated systems, which limits the spatial resolution and makes synchronous /simultaneous measurements challenging. Similarly, temperature sensing with NV and Nickel (Ni) diamond colour centers excited in the biological transparency window in FND crystals was also recently demonstrated[23]. Unfortunately, both the NV and Ni centers show a quite broad phonon sideband, limiting their temperature sensitivity intrinsically[23].

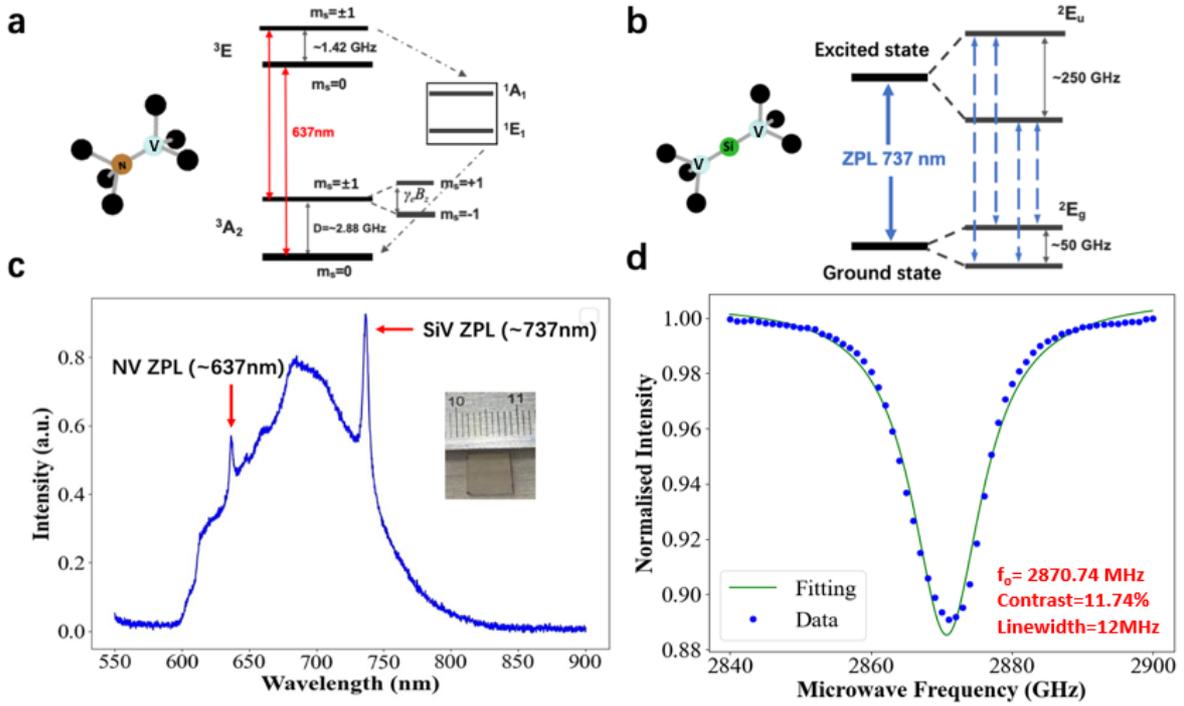

**Figure 1.** (a) Energy level diagram and atomic structure of the NV center. The inset shows the NV center crystal structure (b) Energy level diagram and atomic structure of the SiV defect center. The inset shows the SiV crystal structure (c) PL spectrum of sample at 25 °C, inset is a typical photo of our customized diamond sample. (d) Typical CW-ODMR spectrum at 25 °C with the relevant parameters extracted from Lorentzian Data fitting.)

In this work, we demonstrate a cross-validated/dual-mode optical thermometry method using NV and SiV defect centers in the same bulk diamond sample, without any loss in spatial or temporal resolution. Temperature measurements will be obtained simultaneously using two different modalities, i.e., thermally induced ODMR spectrum and PL spectrum shift, corresponding to NV and SiV centers, respectively. We show that these two measurements have a perfect linear dependency, indicating the confidence of the performed temperature measurement can be significantly improved. Furthermore, we have applied this method to demonstrate reliable thermometry, even in the presence of fluctuating electromagnetic fields which mimics a commonly encountered artefact in practical systems. To the best of our knowledge, this is the first time that people have combined the high sensitivity spin-based approach (NV) with the artefact-free all-optical approach (SiV) within the same thermometer to perform reliable temperature measurements. Specifically, by simultaneously measuring temperature using two different modalities having different physics, our approach avoids sensor confusion and improves the measurement confidence. This could certainly help in addressing the recent controversy surrounding the interpretation of the heterogeneous temperature distribution in living cells[24,25].

## 2. Results

**Co-existence of NV and SiV centers in diamond**

The NV center is a point defect in the diamond lattice, consisting of an adjacent pair of a nitrogen impurity and a lattice vacancy (Inset, **Figure 1**(a)). The electronic energy level structure of the NV centers is shown in Figure 1(a), consisting of spin-1 system (S = 1) with a triplet ground state ($^3A_2$) having electron sublevels of $m_S = 0$ and $m_S = \pm 1$. By applying the microwave resonance to the transition between $m_S = 0$ and $m_S = \pm 1$ in the ground state, the fluorescence decreases substantially—and this process is called ODMR[8]. ODMR measurements can be performed to measure the Zero Field Splitting (D), which is equal to D = 2.87 GHz at room temperature. NV-ODMR based thermometry is based on the temperature dependence of D, originating from thermal expansion of the diamond lattice and the temperature dependence of the electron–phonon interaction [16,26].

The SiV center is formed by replacing two neighbouring carbon atoms in the diamond lattice with one Silicon atom, which places itself between the two vacant lattice sites (Inset, Fig 1(b)). The electronic energy level structure of the SiV centers is shown in Figure 1(b), consisting of doubly degenerate ground and excited orbital states split by spin-orbit coupling[27]. All four transitions between the two ground and two excited orbital states are dipole allowed with a sharp zero phonon line (ZPL) at 738 nm (1.68 eV)[28,29] and a minimal phononic sideband in a roughly 20 nm window around 766 nm[30]. SiV based thermometry is based on the temperature dependence of the ZPL parameters such as peak position and linewidth.

The SiV center emits much more of its emission into its ZPL, approximately 70% (Debye Waller Factor (DWF) =~0.7) as compared to the NV center (0.04)[31]. Moreover, the ZPL is in the Near Infrared Region (NIR) and lies in the biological transparency window. These factors make the SiV center an attractive candidate for a variety of ultrasensitive all-optical thermometry-based applications[32,33].

A detailed optical characterization for our bulk diamond sample using our custom-built wide-field quantum diamond microscope is performed (see Fig. S1, Supporting Information), by measuring the photoluminescence (PL) spectrum and Continuous Wave (CW) - ODMR spectrum to experimentally confirm the coexistence of dual defect centers. Figure 1 (c) shows the PL spectrum measured at 25 degrees Celsius under 532 nm excitation. The ZPL peaks of NV and SiV at 637nm and 737nm respectively can be clearly observed, indicating their coexistence.

By performing the appropriate Lorentzian fitting, the linewidths for the NV and SiV ZPL peaks are extracted as ~3 nm and ~4.8 nm respectively, which are in close agreement to values reported for typical 1b diamond at room temperature[34]. Additionally, a standard CW-ODMR spectrum measurement at 25 degrees Celsius is also performed (Figure 1(d)) which further confirms the presence of an ensemble of NV centers in the bulk diamond sample. A Lorentzian data fitting is performed to extract the ODMR contrast (~12%) and linewidth (12MHz). Interestingly, the ODMR measurement is performed successfully under the influence of the SiV fluorescence as a background signal (with a 650 nm longpass filter).

This demonstrates the mutual independence of the two methods and shows no significant crosstalk between the two defect centers in the diamond sample. This is consistent with recent literate reports about the coexistence of two defect centers in the same bulk diamond sample[35]. A material characterization of the bulk diamond sample used is also provided in the Supporting Information (XRD/Raman measurements are shown in Figure S2), indicating the high crystalline quality of our sample. Additional measurements need to be performed to quantitatively estimate the exact NV/SiV concentrations in the bulk diamond sample.

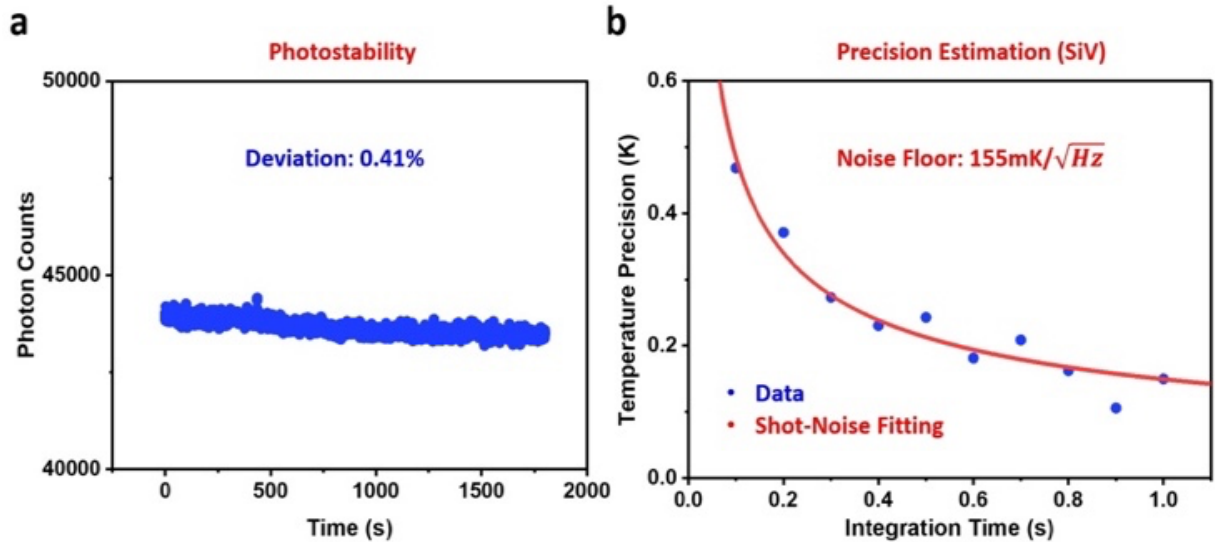

**Figure 2.** (a) The monitored time-trace of fluorescence in diamond sample for a period of 30 minutes. The photon counts are measured for an exposure time of 10 ms and averaged over 400 pixels. (b) The measured temperature precision of the all-optical SiV-based temperature measurement as a function of integration time. A noise floor of 155 mK/√Hz is extracted from the shot-noise fitting.

Apart from the PL and ODMR measurements, the performance of the 2 thermometry methods also needs to be characterized, to confirm whether the co-existence of the two defect centers in the sample has any adverse effects on the sensitivity. The photostability of the sample was measured by recording photon counts under 532nm excitation for over 30 minutes at 25 degrees Celsius, and a deviation of less than 0.41% was measured (**Figure 2(a)**). To estimate the precision of SiV thermometry, we measure the uncertainty (standard deviation) in the ZPL peak position as a function of integration time at a fixed temperature of 25 degrees Celsius (Figure 2 (b)). By performing the appropriate shot-noise limited data-fitting[18] to the obtained curve, a thermal noise floor of 155 mK/√Hz can be extracted. Similarly, to estimate the precision of NV-ODMR based thermometry, a Lorentzian Data-fitting on the CW-ODMR spectrum is performed to extract the contrast and linewidth, and then the well-known theoretical relation[8] is used to calculate the sensitivity of 22 mK/√Hz (see Supporting Information). The measured noise floors for both methods are comparable to the sensitivities reported in previous works[8].

**Cross-validated Temperature Measurement**

Temperature measurements are demonstrated by measuring the NV-ODMR resonance frequency, SiV ZPL parameters (Peak Position, Peak FWHM) as a function of diamond sample temperature. At each temperature, the ODMR and PL Spectrum measurements are performed, and the parameters mentioned above are extracted by performing the relevant Data fitting. The temperature of the diamond sample was stabilized using an electronic temperature-controlled system (global heating source), having a precision of 0.1 °C.

**Figure 3**(a) demonstrates NV-ODMR based thermometry where the resonance frequency is measured as a function of temperature, and a strong linear dependence is observed. A susceptibility of 73.79 kHz/°C is derived from linear fitting, which closely matched previously reported values[8]. Similarly, Figure 3(b) demonstrates all-optical SiV-based temperature measurement where the ZPL Peak position and FWHM are shown to have a strong linear dependence on temperature, and the susceptibilities are derived from the corresponding linear fitting. The key result of our project is shown in Figure 3 (c), where the temperature measurement is performed simultaneously using the NV-ODMR and SiV-ZPL shift-based methods. The accurate linear relationship (high R2 value) between the ODMR resonance frequency and SiV ZPL position provides strong evidence that the two independent mechanisms can be used to cross-validate each other simultaneously and synchronously.

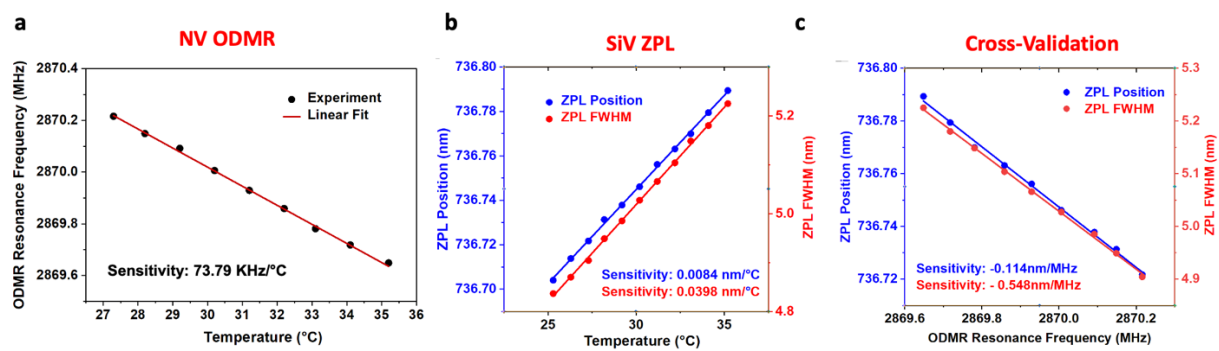

**Figure 3**. (a) ODMR Resonance Frequency is measured as a function of temperature, and a thermal susceptibility of 73.79 kHz/°C is extracted. Error bars represent the standard deviation calculated from 10 individual measurements. (b) The SiV ZPL position and FWHM are measured as a function of temperatures, and thermal susceptibilities of 0.0084 nm/°C and 0.0398 nm/°C are extracted respectively. (c) The SiV ZPL Position/FWHM measurement is shown as a function of NV-ODMR resonance frequency. Error bars for (a), (b) and (c) are smaller than the size of the datapoint.

Additionally, we further demonstrate cross validation temperature measurements by using the excitation laser (532nm) as a local heating source. The NV ODMR resonance frequency and SiV ZPL peak position are measured simultaneously as a function of laser power, in both the static and dynamic settings (Figure S3). A close agreement between the 2 temperature measurement methods is found, which practically demonstrates how the all-optical SiV ZPL measurements can be used to verify the spin-based NV ODMR measurements in a stable and repeatable manner. (See Supporting Information for experimental details). After demonstrating the viability and feasibility of our cross-validation approach, we explore a practical application of this method in the next section.

**Practical application of the developed cross-validated method**

The main driving force of this project is that the NV-ODMR measurement is susceptible to multiple artefacts such as external fluctuating electromagnetic fields, microwave induced heating, etc. This hampers the quality of the measurements and inevitably reduces the accuracy and reliability of the measurement performed. As a comparison, the SiV-based temperature method is all-optical, and is immune to the above-mentioned artefacts.

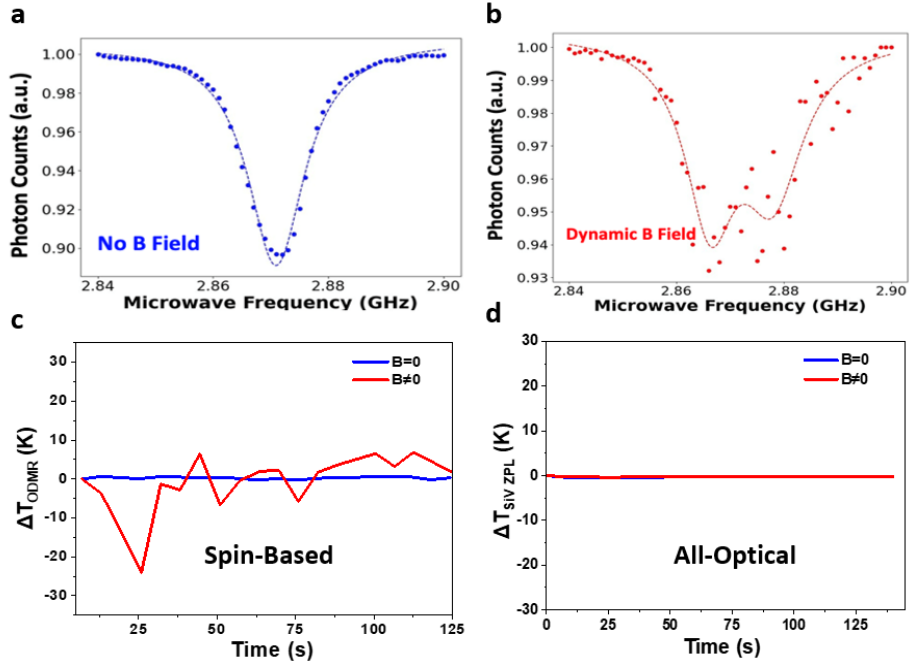

**Figure 4**. (a) Influence of dynamic magnetic fields on the ODMR-based diamond thermometry. (a) CW-ODMR spectra shown absence of B field. (b) CW-ODMR spectra in the presence of randomly oriented dynamic B field (c) Time trace of measured temperature change for NV-ODMR based method. (d) Time-trace of measured temperature change for all-optical SiV ZPL based method.

As a proof-of-concept demonstration, we study the effect of a randomly oriented dynamic magnetic field as a measurement artefact for NV ODMR-based measurement. The applied B field is applied manually by the experimenter and fluctuates randomly both in magnitude and direction, with a typical frequency on the order of Hz. This mimics a common measurement artefact (fluctuating electromagnetic fields) found in many systems, e.g., in complex biological environments such as neuron cells and cardiac tissues[36]. While it is possible to accurately measure temperature in the presence of magnetic noise (and vice versa) either by using multi-point methods[6] or by performing a mathematical analysis on the fitted resonance frequencies[8,9], these methods only work if the B field can be assumed to be static during the measurement time of the ODMR spectrum.

Figure 4 ((a) and (b)) demonstrates how an external magnetic field leads to the splitting of the ODMR spectrum, resulting in a lower contrast and a higher linewidth. These factors lead to a degraded temperature sensitivity for the NV based ODMR measurement. As a proof-of-concept demonstration, we measure the ODMR resonance frequency as a function of time (Figure **4**(c)), both in the presence and absence of an external fluctuating magnetic field. As expected, the measured temperature change has a large variance and standard deviation when a dynamic external magnetic field is applied. On the contrary, when the same temperature

measurement is performed for the all-optical SiV-based spectrum parameters (Figure **4**(d)), there is comparably a significantly lower variance when an external magnetic field is applied. This demonstrates how the SiV-based method can be used to validate the NV-based ODMR method in the presence of measurement artefacts, as shown for fluctuating magnetic fields.

## 3. Conclusion

In this letter, we have proposed that it is feasible to effectively combine the high sensitivity, accuracy, and stability of NV-ODMR measurement, with the advantages of artefact-free all-optical SiV approach, without any loss in spatial or temporal resolution, to improve the reliability and measurement confidence of temperature measurements. We envision a variety of novel applications in the future by combining multiple defect centers in the same diamond sample, simultaneously utilizing the advantages and use-cases of different colour centers together.

For example, dual-defect centers can allow us to decouple the Temperature (T) and B field measurements, significantly simplifying the measurement and making it more reliable. Simultaneous T and B measurements have multiple applications and have performed multiples times before [8,9]. However, this involves mathematical calculations on the multiple extracted resonance frequencies and can only be accurately and reliably done when T and B are static during the time taken to perform the measurement. Using dual-mode quantum sensing, i.e., SiV for T measurements and NV for B-field measurement, allows us to make precise magnetic field measurements in the presence of temperature fluctuations and vice versa.

Furthermore, while the current measurements were performed on a bulk diamond sample which might limit the scope of its practical applications, we propose a sample fabrication procedure to obtain nanodiamonds containing dual defect centers. That would allow cross-validation of temperature measurement with a nanoscale spatial resolution. The nanodiamond sample fabrication procedure has been described in the Supporting Information, and preliminary experimental results are shown in Figure S4. The detection of simultaneous NV and SiV ZPL peaks in the PL Spectrum measurement indicates the co-existence of both defect centers in a single nanodiamond. Further sample optimisation is required in order to improve the SNR of the ZPL peaks, following which a detailed characterisation of the sensitivity and performance for the nanodiamond case can be performed. Moreover, performing a ball milling process on the currently used bulk diamond sample[37] shall also yield nanodiamonds having a coexistence of NV and SiV.

In conclusion, in this work, we have succeeded in measuring temperature simultaneously using 2 independent mechanisms (NV-ODMR and SiV-ZPL) without any substantial loss in sensitivity or temporal resolution. The use of two modalities enables cross-validation and makes these results far more reliable for nanothermometers in complex systems such as living cells.

**Acknowledgements**

Z.Q.C. acknowledges the financial support from Hong Kong SAR Research Grants Council (RGC) Early Career Scheme (ECS; No. 27202919); Hong Kong SAR RGC Research Matching Grant Scheme (RMGS; No. 207300313); Hong Kong SAR Innovation and Technology Fund (ITF) through the Platform Projects of the Innovation and Technology Support Program (ITSP; No. ITS/293/19FP); and HKU Seed Fund (No. 202011159019 and No. 202010160007).

**Supporting Information**

Supporting Information is available upon request from the author.

**Conflicts of interest**

There are no conflicts to declare.

**Data Availability Statement**

The data that support the findings of this study are available from the corresponding author upon reasonable request.

**References**

[1]    T. Zhang, G. Pramanik, K. Zhang, M. Gulka, L. Wang, J. Jing, F. Xu, Z. Li, Q. Wei, P. Cigler and Z. Chu, *ACS Sens*ors, 2021, **6**, 2077−2107.

[2]    R. Schirhagl, K. Chang, M. Loretz and C. L. Degen, *Annual Review of Physical Chemistry.*, 2014, **65**, 83−105.

[3]    M. W. Doherty, N. B. Manson, P. Delaney, F. Jelezko, J. Wrachtrup and L. C. Hollenberg, *Physics Rep*orts, 2013, **528**, 1−45.

[4]    T. Muller, C. Hepp, B. Pingault, E. Neu, S. Gsell, M. Schreck, H. Sternschulte, D. Steinmuller-Nethl, C. Becher and M. Atature, *Nature Communications*, 2014, **5**, 3328.

[5]    Herbert Precht, Jes Christophersen , Herbert Hensel , Walter Larcher, Temperature and Life, 1973  (Berlin: Springer)

[6]     Zhou, J., del Rosal, B., Jaque, D., Uchiyama, S., Jin D., *Nature Methods,* 2020,**17**, 967–980.

[7]     Kucsko, G., Maurer, P., Yao, N., Kubo, M., Noh, H., Lo, P., Park, H., Lukin, M., *Nature*, 2013, **500**, 54–58.

[8]     Masazumi Fujiwara and Yutaka Shikano, *Nanotechnology*, 2021, **32**, 482002

[9]     Christopher Foy, Lenan Zhang, Matthew E. Trusheim, Kevin R. Bagnall, Michael Walsh, Evelyn N. Wang, and Dirk R. Englund, *ACS Applied Materials Interfaces*, 2020, **12(23)**, 26525–26533

[10]    Fujiwara M, Sun S, Dohms A, Nishimura Y, Suto K, Takezawa Y, Oshimi K, Zhao L, Sadzak N, Umehara Y, Teki Y, Komatsu N, Benson O, Shikano Y, Kage-Nakadai E. Real-time nanodiamond thermometry probing in vivo thermogenic responses. Sci Adv. 2020 Sep 11;6(37):eaba9636. doi: 10.1126/sciadv.aba9636. PMID: 32917703; PMCID: PMC7486095

[11]    Hannah Clevenson, Edward H. Chen, Florian Dolde, Carson Teale, Dirk Englund, and Danielle Braje, *Phys. Rev. A*, 2016, 94 021401

[12]    Paolo Andrich, Jiajing Li, Xiaoying Liu, F. Joseph Heremans, Paul F. Nealey, and David D. Awschalom, Nano Letters, 2018, **18**, 4684–90

[13]    Zheng Wang, Jintao Zhang, Xiaojuan Feng, Li Xing, *ACS Omega*, 2022, **7(35)**, 31538–31543

[14]    Yushi Nishimura, Keisuke Oshimi, Yumi Umehara, Yuka Kumon, Kazu Miyaji, Hiroshi Yukawa, Yutaka Shikano, Tsutomu Matsubara, Masazumi Fujiwara, Yoshinobu Baba & Yoshio Teki, *Scientific Reports*, 2021, **11**, 4248

[15]    Nguyen J P, Shipley F B, Linder A N, Plummer G S, Liu M, Setru S U, Shaevitz J W, Leifer A M, *Proceedings of the National Academy of Sciences of the United States of America*, 2016, **113** E1074–81

[16]    Taras Plakhotnik, Marcus W. Doherty, Jared H. Cole, Robert Chapman, Neil B. Manson, *Nano Letters*, 2014, **14(9)**, 4989–4996

[17]    Choi, S.; Agafonov, V.; Davydov, V.; Plakhotnik, T. *ACS Photonics*, 2019, **6**, 1387

[18]    Christian T. Nguyen, Ruffin E. Evans, Alp Sipahigil, Mihir K. Bhaskar, Denis D. Sukachev, Viatcheslav N. Agafonov, Valery A. Davydov, Liudmila F. Kulikova, Fedor Jelezko, and Mikhail D. Lukin*, Applied Physics Letters*, 2018, **112(20)**, 203102

[19]    Jing-Wei Fan, Ivan Cojocaru, Joe Becker, Ilya V. Fedotov, Masfer Hassan A. Alkahtani, Abdulrahman Alajlan, Sean Blakley, Mohammadreza Rezaee, Anna Lyamkina, Yuri N. Palyanov, Yuri M. Borzdov, Ya-Ping Yang, Aleksei Zheltikov, Philip Hemmer, and Alexey V. Akimov, *ACS Photonics*, 2018, **5, 3**, 765–770

[20]   S. Ditalia Tchernij, T. Lühmann, T. Herzig, J. Küpper, A. Damin, S. Santonocito, M. Signorile, P. Traina, E. Moreva, F. Celegato, S. Pezzagna, I. P. Degiovanni, P. Olivero, M. Jaksič, J. Meijer, P. M. Genovese, J. Forneris, *ACS Photonics*, 2018, **5(12)**, 4864–4871

[21]   Masfer Alkahtani, Ivan Cojocaru, Xiaohan Liu, Tobias Herzig, Jan Meijer, Johannes Kupper, Tobias Luhmann, Alexey V. Akimov, Philip R. Hemmer, *Applied Physics Letters*, 2018, **112**, 241902.

[22]   Masfer Alkahtani, Linkun Jiang, Robert Brick, Philip Hemmer, and Marlan Scully, Optics Letters, 2017, **42**, 4812-4815

[23]   Masfer H. Alkahtani, Fahad Alghannam, Linkun Jiang, Arfaan A. Rampersaud, Robert Brick, Carmen L. Gomes, Marlan O. Scully, Philip R. Hemmer, Optics Letters, 2018, **43**, 3317-3320.

[24]   Suzuki, M., Plakhotnik, T. The challenge of intracellular temperature. *Biophys Rev* **12**, 593–600 (2020). https://doi.org/10.1007/s12551-020-00683-8

[25]   Sushrut Ghonge and Dervis Can Vural, Temperature as a quantum observable, *Journal of Statistical Mechanics,* (2018) 073102

[26]   X.-D. Chen, C.-H. Dong, F.-W. Suna), C.-L. Zou, J.-M. Cui, Z.-F. Han, and G.-C. Guo, Applied Physics Letters, 2011, **99**, 161903

[27]   Benjamin Pingault, David-Dominik Jarausch, Christian Hepp, Lina Klintberg, Jonas N. Becker, Matthew Markham, Christoph Becher & Mete Atatüre, *Nature Communications*, 2017, **8**, 15579

[28]   Y. Liu, G. Chen, Y. Rong, L. P. McGuinness, F. Jelezko, S. Tamura, T. Tanii, T. Teraji, S. Onoda, T. Ohshima, J. Isoya, T. Shinada, E. Wu and H. Zeng, *Scientific Reports*, 2015, **5**, 12244.

[29]   Tom Feng, Bradley D. Schwartz, *Journal of Applied Physics*, 1993, **73 (3)**:1415.

[30]   Dietrich, A.; Jahnke, K. D.; Binder, J. M.; Teraji, T.; Isoya, J.; Rogers, L. J.; Jelezko, F. *New Journal of Physics*, 2014, **16 (11):** 113019.

[31]   I. Aharonovich, S. Castelletto, D.A. Simpson, C-H Su, A. D. Greentree, S Prawer, Reports on Progress in Physics, 2011, 74(7), 076501

[32]   Tongtong Zhang, Madhav Gupta, Jixiang Jing, Zhongqiang Wang, Xuyun Guo, Ye Zhu, Yau Chuen Yiu, Tony K.C. Hui, Qi Wang, Kwai Hei Li, Zhiqin Chu, *Journal of Material Chemistry C*, 2022, **10**, 13734

[33]   Weina Liu, Md Noor A. Alam, Yan Liu, Viatcheslav N. Agafonov, Haoyuan Qi, Kaloian Koynov, Valery A. Davydov, Rustem Uzbekov, Ute Kaiser, Theo Lasser, Fedor Jelezko, Anna Ermakova, Tanja Weil, *Nano Letters,* 2022, **22(7)**, 2881–2888

[34] B. Dong, C. Shi, Z. Xu, K. Wang, H. Luo, F. Sun, P. Wang, E. Wu, K. Zhang, J. Liu, Y. Song, Ye Fan, *Diamond Related Materials*, 2021, 108389

[35] Lena Golubewa, Yaraslau Padrez, Sergei Malykhin, Tatsiana Kulahava, Ekaterina Shamova, Igor Timoshchenko, Marius Franckevicius, Algirdas Selskis, Renata Karpicz, Alexander Obraztsov, Yuri Svirko, Polina Kuzhir, *Advanced Optical Materials*, 2022,**10**, 2270060.

[36] Keigo Arai, Akihiro Kuwahata, Daisuke Nishitani, Ikuya Fujisaki, Ryoma Matsuki, Yuki Nishio, Zonghao Xin, Xinyu Cao, Yuji Hatano, Shinobu Onoda, Chikara Shinei, Masashi Miyakawa, Takashi Taniguchi, Masatoshi Yamazaki, Tokuyuki Teraji, Takeshi Ohshima, Mutsuko Hatano, Masaki Sekino & Takayuki Iwasaki, *Communications Physics*, 2022,**5**,200 (2022)

[37] Lin, C.R., Wei, D.H., Dao, M.K.B., Chung, R.J., Chang, M.H., 2013. Nanocrystalline Diamond Particles Prepared by High-Energy Ball Milling Method. AMM 284–287, 168–172. https://doi.org/10.4028/www.scientific.net/amm.284-287.168

# Supporting Information

**Cross-correlated quantum thermometry using diamond containing dual-defect centers**

*Madhav Gupta[a], Tongtong Zhang[a], Lambert Yeung[b], Jiahua Zhang[a], Yayin Tan[a], Yau Chuen Yiu[a], Shuxiang Zhang[a,e], Qi Wang[c], Zhongqiang Wang[c], Zhiqin Chu*[a,d]*

[a] Department of Electrical and Electronic Engineering, The University of Hong Kong, Pokfulam Road, Hong Kong SAR, China
[b] 6C Tech Ltd. Hong Kong SAR, China
[c] Dongguan Institute of Opto-Electronics, Peking University, Guangdong, China
[d] School of Biomedical Sciences, The University of Hong Kong, Pokfulam Road, Hong Kong SAR, China
[e] Sichuan University, China

E-mail: zqchu@eee.hku.hk

## Experimental Section

### Sample Fabrication

#### Bulk Diamond

The growth of diamond substrate samples was performed in a metallic Microwave Plasma-Chemical Vapor Deposition (MPCVD) reactor developed by 6C Technology Limited for growing monocrystalline diamond in a resonant cavity at 915MHz. Synthetic (100) chemical-vapor-deposition (CVD) diamonds were used as substates for homoepitaxial diamond growth. These crystals were carefully cleaned in boiling sulfuric acid and nitric acid to remove any contaminants such as metals or graphite. After the substrates were placed in the chamber was evacuated with a roughing pump until a pressure under 1E-2 was reached. The CVD diamond deposition procedure consisted of a two-step process. The first was a 10~30 min pre-treatment with a hydrogen plasma. The second step was the actual diamond growth, which consisted in a plasma deposition ignited with a microwave power of 10~20kW at a total pressure of 10~20kPa. The substrate temperature was varied in the range 850–1000 C, and the methane concentration was varied from 5% to 7% in respect to hydrogen flow of 600 sccm.

#### Nanodiamond (Figure S4)

Salt-assisted air oxidation (SAAO) of NDs: 0.5 g NDs with a mean particle size of 50 nm (HPHT, PolyQolor, China) were mixed with 2.5 g Sodium Chloride (NaCl, 99.5 %, Sigma-Aldrich), and they were heated at 500 °C for 1 hour in air. The resultant sample was dispersed in deionized (DI) water and sonicated for 1 hour, and the NDs were then purified with deionized (DI) water three times by centrifugation. The purified NDs were re-dispersed in DI water and sonicated for 10 min to obtain well-dispersed NDs suspension for the CVD diamond growth.

Chemical vapor deposition (CVD) of diamond microparticles on Si wafer: The Si substrate was firstly treated with hydrogen plasma by the microwave-plasma assisted chemical vapor deposition (MPCVD) system (Seki 6350) for 10 minutes. Then the CVD diamond seeds, i.e., 50 nm HPHT SAAO NDs and commercially available nanodiamonds (DNDs, <10nm, TCI) were spin-coated on the standard single-crystal Si (100) wafers (2 inches). And the diamond films were grown with a gas mixture of $H_2/CH_4$ (94/6) under fixed power and pressure conditions for 80 minutes.

Optical Characterization

The experiments were performed on the well-studied quantum diamond widefield microscope, shown in Figure S1 A 532nm laser source is used to excite the diamond sample in a widefield configuration using a high NA objective (MPLAPON60x). A high sensitivity EMCCD camera (Andor iXon 897) is used to measure the photon counts, whereas a Thorlabs spectrometer (CCS175/M) is used to perform the spectrum measurements. The microwave system generates the required microwave signals to perform the required ODMR measurements, whereas the Pulse Generator is used to perform synchronization between different components. The setup allows us to perform the 2 key measurements required for this protocol: 1) Continuous Wave (CW)-ODMR and 2) Photoluminescence (PL) Spectrum. The temperature of the diamond sample was stabilized using an electronic temperature-controlled system. The sample was glued to a metal ceramic heater (Thorlabs HT24S) and a platinum thermistor (Thorlabs, TH100PT). The heater and the thermistor were both electronically connected to a proportional-integral derivative (PID) temperature controller (Thorlabs, TC200), stabilizing the sample temperature with a precision of ±0.1 °C

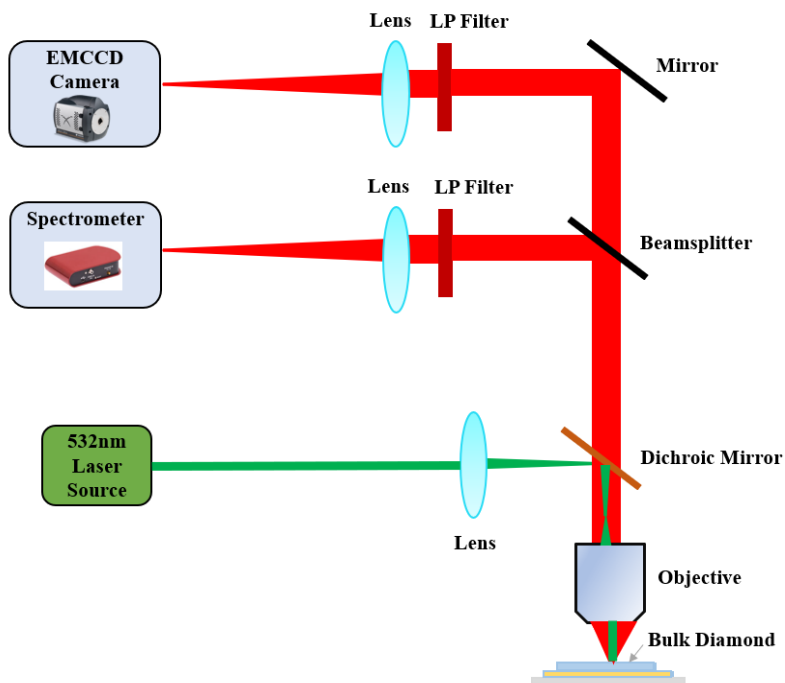

**Figure S1:** Schematic showing the optical setup for the widefield quantum diamond microscope used for the measurements

## Material Characterisation

The X-ray diffraction (XRD) results (Fig. S2(a)) indicate the pure crystalline nature of the bulk diamond sample, i.e., only the characteristic peaks of the diamond (111) plane are found in the XRD spectra. Raman analysis as shown in Fig. S2(b) presents a well-defined diamond Raman peak at 1333 cm−1 of the bulk diamond sample. However, some non-diamond impurities (1450 ∼ 1550 cm−1) are observed as well, apart from the diamond peak.

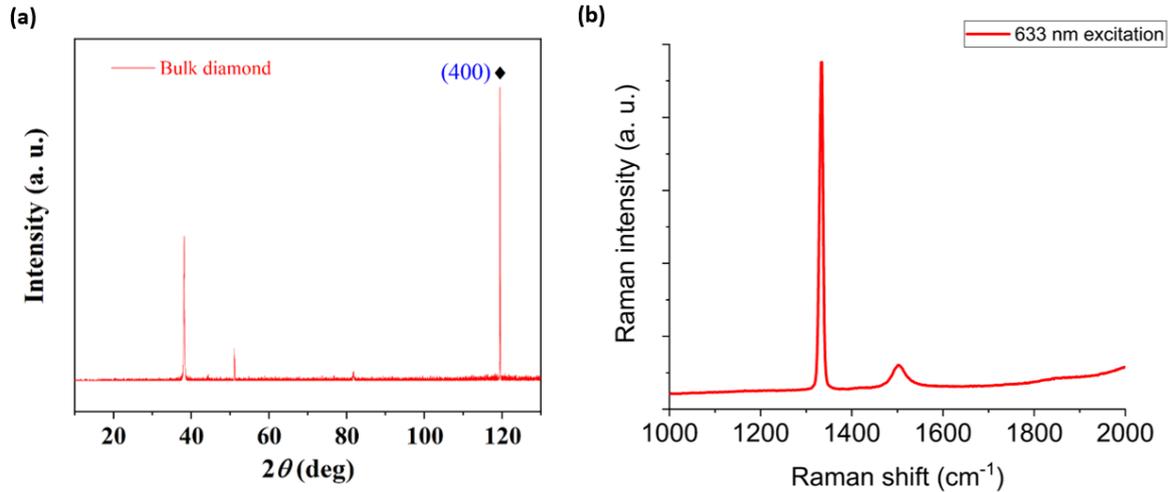

**Figure S2** (a) XRD and (b) Raman measurement of the bulk diamond sample containing dual defect centers (NV & SiV)

## Photostability and Precision Estimation

The photon counts are measured using an Andor iXon 897 camera for an ROI of 6um*6um (averaged over 400 pixels), at a fixed laser power and temperature (25 degrees Celsius) for a period of 30 minutes, and a deviation of only 0.41% is measured (Figure S2(a)).

To calculate the thermal noise floor, the standard deviation of SiV ZPL peak position is measured as a function of different integration times of the spectrometer. The measured standard deviation is divided by the measured thermal susceptibility (Figure 2 (b) of manuscript) to calculate the temperature precision. The appropriate shot-noise limited fitting is applied to the obtained curve to get the thermal noise floor η = 155mK (Figure S2 (b)). This measurement uncertainty follows the shot-noise limit $1/\sqrt{t}$ (Datafitting), indicating that the temperature precision can be further improved by increasing photon collection rates from the sample, either by increasing SiV density or by improving the collection efficiency.

To estimate the precision of NV-ODMR based thermometry, we perform a Lorentzian Data-fitting on the CW-ODMR spectrum to extract the contrast and linewidth, and then the following well-known theoretical relation [7] is used to calculate the sensitivity: $\eta=\Delta w/(C\sqrt{R}|dD/dT|)$, where C, Δω, R, and dD/dT denote the ODMR contrast, ODMR linewidth, detected photon rates, and temperature dependence of the ZFS, respectively. By substituting C=0.12, Δω=12, dD/dT =0.074, R= 10 Mcps, we get η=22mK.

Using these calculation/measurement methods a thermal noise floor of 22mK/√Hz for the NV-ODMR based measurement, and 155mK/√Hz for the all-optical SiV-based measurement was calculated, which are close to the sensitivities reported in previous works [7].

**Synchronised Cross-Validation using Local Heating Source (Laser Power):**

Bulk diamond sample having a co-existence of NV and SiV centers is used, with a layer of gold nanodiamonds deposited on its surface. The gold nanodiamonds allow local heating of the diamond surface, with the temperature being proportional to the applied laser power. The laser power is modulated by using an Acousto-Optic Modulator. A high temporal resolution pulse generator (Swabian Pulse Streamer) is used to synchronise the ODMR measurement, Spectrum measurement, and laser power modulation in real-time.

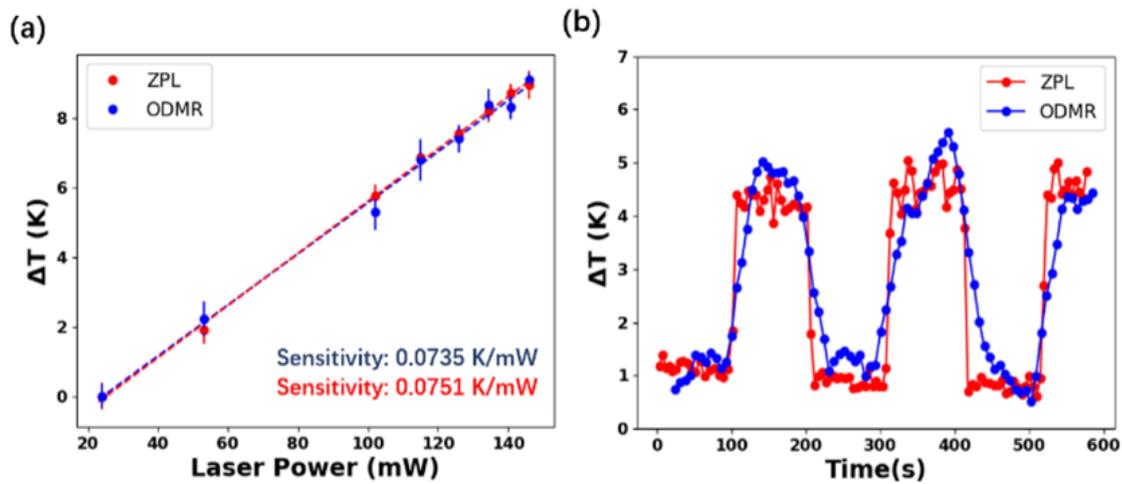

**Figure S3**. (a) ODMR resonance frequency and ZPL position are measured simultaneously as a function of laser power, and a thermal sensitivity of 0.0735 K/mW and 0.0751 K/mW is extracted respectively. The error bars represent the standard calculated from 10 individual measurements. (b) The laser power is periodically modulated in a step like fashion from 85 mW to 145 mW, with a period of ~200 seconds. A stable, repeatable cross validation of temperature measurement is shown in a dynamic setting. The measurement time for each data point is ~1.5 seconds for the ODMR measurement, and ~1.3s for the spectrum measurement.

**Nanodiamond Sample Measurement:**

Preliminary measurements are performed to demonstrate the co-existence of NV and SiV defect centers in a nanodiamond sample. The fabrication procedure for the sample has been described in the experimental section above. A home-built confocal microscope is used to perform these measurements. Fig S4(a) shows a two-dimensional scan of nanodiamonds performed over an ROI of ~80 um, where dispersed particles can be seen. Fig S4(b) shows the SEM image of this sample.

Firstly, a widefield CW- ODMR measurement is performed to confirm the presence of NV centers in the nanodiamond sample (Fig S4(c)). This is followed by a series of PL spectrum measurements on single particles. A large proportion of the particles show characterestic spectra of either only NV (Fig S4(d)) or SiV ((Fig S4(e))). However, a small proportion of particles show a co-existence of both the defect centers, with clear NV & SiV ZPL peaks at 637 nm and 737 nm respectively ((Fig S4(f))).

Further sample optimisation is required to improve the SNR of the ZPL peaks in the PL spectrum, in order to perform nanoscale resolution thermometry.

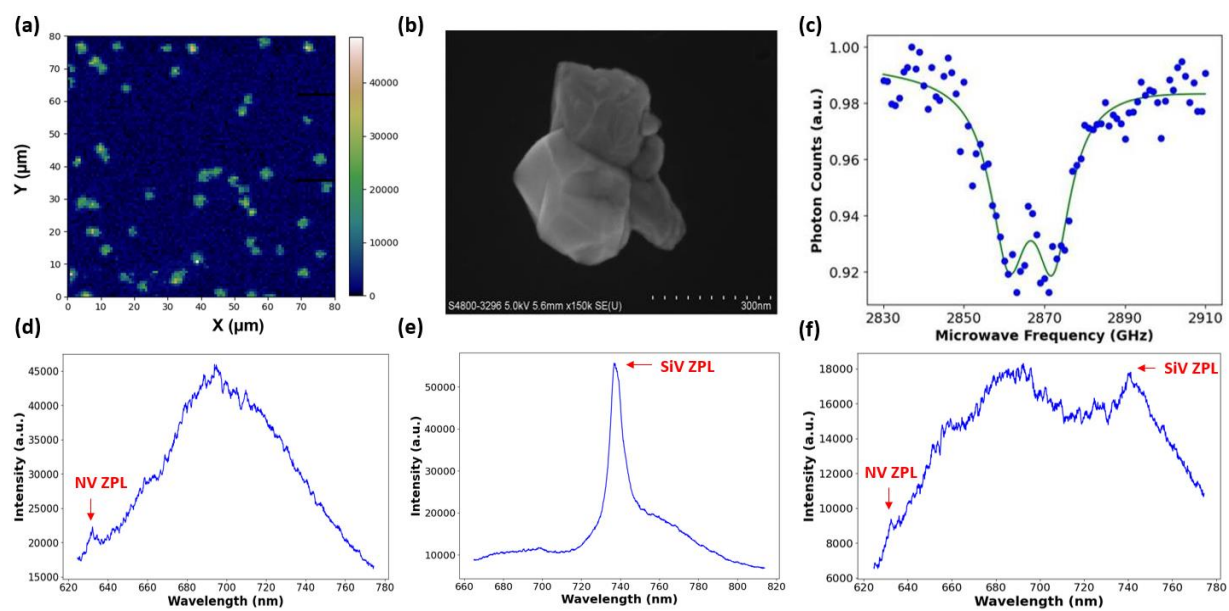

**Figure S4** Co-existence of NV and SiV defect centers in a nanodiamond sample. (a) Two-dimensional confocal scan of nanodiamonds. (b) SEM image of nanodiamond sample. (c) CW-ODMR spectrum measured at room temperature. (d) PL Spectrum of single ND showing clear NV center ZPL peak and it's characteristic phonon sideband (e) PL Spectrum of single ND showing clear SiV ZPL peak (f) PL Spectrum of single ND showing both NV & SiV ZPL peaks, confirming the co-existence of both the defect centers.